\documentclass[prb,twocolumn,aps,floatfix]{revtex4}
\usepackage{graphicx}
\usepackage{bm}
\usepackage{amssymb}
\usepackage{hyperref}
\usepackage{amsmath}
\usepackage{subfigure}
\usepackage{tikz}
\usetikzlibrary{shapes,arrows}
\usepackage{amsmath}
\usepackage{appendix}
\usepackage{multirow}
\usepackage{booktabs}
\usepackage{tikz}

\newcommand{\tabf}{\hspace*{0.2em}}

\begin{document}
\title{Saturation field entropies of antiferromagnetic Ising model: Ladders and the kagome lattice}
\author{Vipin Kerala Varma}
\affiliation{Bethe Center for Theoretical Physics, Universit\"{a}t Bonn, Germany\\}
\date{\today}
\begin{abstract}
 Saturation field entropies of antiferromagnetic Ising models on quasi one-dimensional lattices (ladders) and the kagome lattice 
 are calculated. The former is evaluated exactly by constructing the corresponding transfer matrices, while the latter calculation uses 
 Binder's algorithm for efficiently and exactly computing the partition function of over 1300 spins to give 
 $S_{\textrm{kag}}/k_B = 0.393589(6)$. We comment on the relation of the kagome lattice to the experimental situation in 
 the spin-ice compound Dy$_{2}$Ti$_{2}$O$_{7}$.
\end{abstract}
\maketitle
\section{Introduction}
\begin{figure}[bbp]
\centering
\subfigure[]{\label{ladders}
\includegraphics*[width=4.2cm]{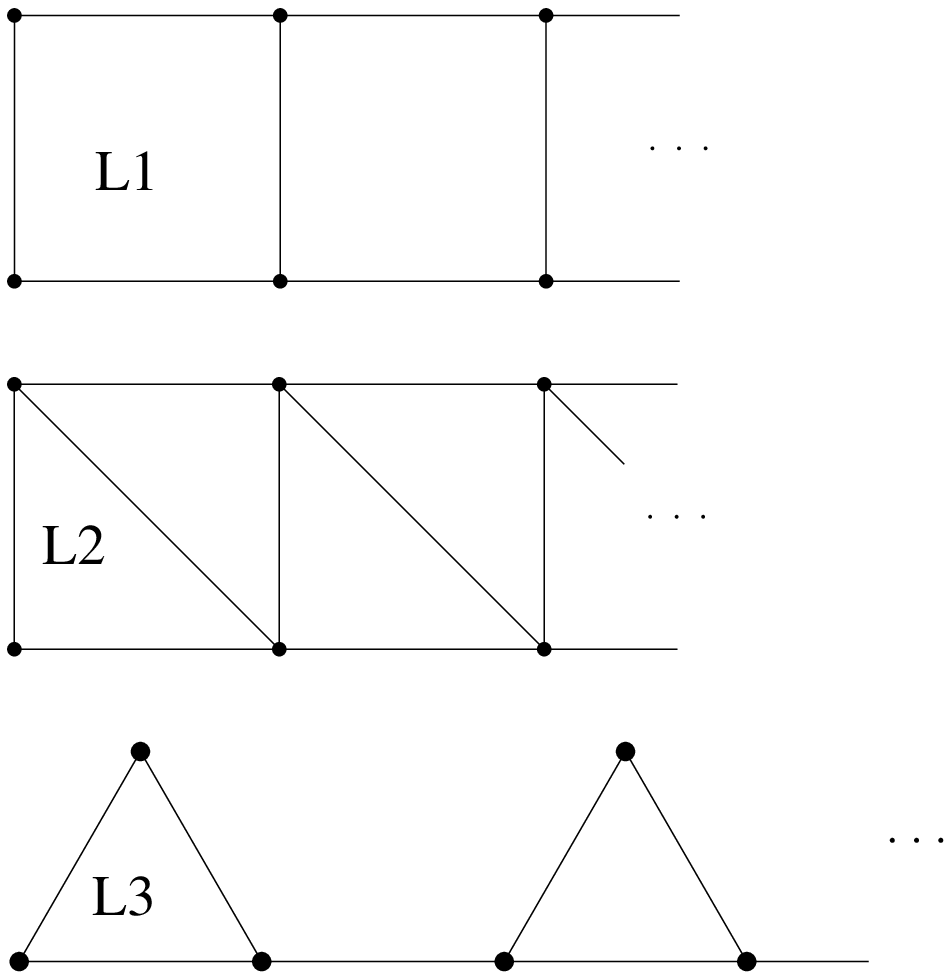}
}\tabf
\subfigure[]{\label{kagome}
 \includegraphics[width=4.6cm]{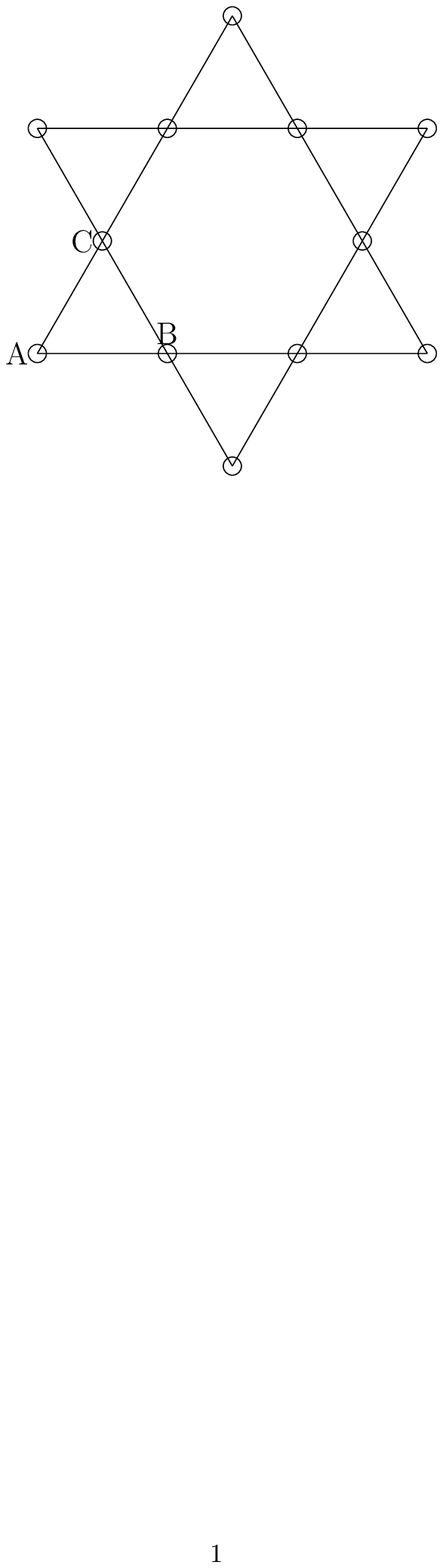}
}
\caption{(a) Quasi one-dimensional lattices (ladders) $L_1, L_2, L_3$. $L_3$ in a triangular lattice pattern reproduces the 
kagome lattice. (b) Kagome lattice with unit cell labelled A,B,C. }
\label{lattice}
\end{figure}
Ising lattice models are realized in a variety of materials with magnetic interactions, for instance in rare earth compounds where the outer electron in the lanthanide ion acts as an Ising spin interacting with its 
neighbors. Depending on the lattice structure (frustrated or unfrustrated),
dimensionality (one to three), and interaction type (ferromagnetic or antiferromagnetic), the properties and phases in the system 
display a wide range of possibilities \cite{Liebmann}. Coupling of the Ising spins to an external field can further introduce 
interesting behavior such as the presence of a magnetization plateau. Many such characteristics of the system may be captured by
the model's partition function; accurate determination of the partition function poses a numerical challenge if there are 
competing interactions within the system, which is often the case for antiferromagnetic interactions on non-bipartite lattices. 
These competing interactions can result in large degeneracies in phase space.\\
Indeed, antiferromagnetic Ising models can harbor a macroscopic number of degenerate ground states at zero external field on 
frustrated lattices like the triangular lattice \cite{Wannier}, kagome lattice \cite{Kano}, pyrochlore \cite{Anderson, Liebmann} 
lattice, to name a few. An extensive entropy may survive, albeit with different values, even for infinitesimal fields 
\cite{Udagawa, Moessner, Isakov} on certain lattices; as the field is varied, a strongly enhanced peak in the entropy develops 
just before the field-induced spin-ordering sets in \cite{Domb, Metcalf, Isakov}; 
this substantial peak occurs because, at this field strength, a large number of non-neighboring spins may be flipped against 
the field without a cost in energy \cite{Metcalf}.
In fact, such residual saturation entropies $S^{\textrm{sat.}}$ persist in quantum spin models (anisotropic Heisenberg models), 
although with different values from the Ising limits, and for different reasons \cite{Schulenberg, Derzhko} pertaining to the 
existence of localized magnons.\\
In section \ref{sec: ladders}, residual saturation entropies of related Ising quasi one-dimensional lattices or 
ladders (Fig. \ref{ladders}) are exactly computed. The statistical properties of the ladders are considered primarily to establish the results on the kagome lattice; 
$L_1$ will be used to construct the square lattice to compare with earlier results before we employ ladder $L_3$ to build up the kagome lattice. The bounds on the 
kagome lattice saturation entropy will be justified via the construction of ladder $L_2$ from $L_1$.
Moreover, ladders with these geometries are realized in a range of materials \cite{DagottoRice, Ohwada, Heuvel}.
The two-leg ladder $L_1$ with magnetic interactions are realized in compounds such as vanadyl pyrophosphate $\mathrm{(VO)}_2\mathrm{P}_2\mathrm{O}_7$ and 
the cuprate $\mathrm{SrCu}_2\mathrm{O}_7$ \cite{DagottoRice}; ladder $L_2$, which is equivalent to the axial-next-nearest-neighbor Ising
(ANNI) model, describes phase transitions in the charge-lattice-spin coupled system $\mathrm{NaV}_2\mathrm{O}_5$ \cite{Ohwada}. 
And ladder $L_3$ bears resemblance in geometric structure and interaction to the Ising-Heisenberg polymer chain 
$\mathrm{[DyCuMoCu]}_{\infty}$ \cite{Heuvel}. \\
In sections \ref{sec: kagome} and \ref{sec: Binders}, we consider the Ising kagome lattice, shown in Fig. \ref{kagome}, at saturation, which was 
argued to be realized in the spin-ice compound Dy$_{2}$Ti$_{2}$O$_{7}$ \cite{Isakov}. 
For the kagome lattice, approximate values of $S^{\textrm{sat.}}$ may be deduced from calculations for spin ice on the pyrochlore 
lattice in a [111] field \cite{Isakov}; results of Monte Carlo simulations, series expansion techniques \cite{Nagle} and the Bethe 
approximation were found to be comparable \cite{Isakov} for the saturation entropy.
In section \ref{sec: kagome}, we elucidate a procedure for obtaining a more accurate estimate of this value through (a) transfer matrix methods, 
and equivalently (b) the solution of appropriate difference equations that generate the partition function. 
Finally in section \ref{sec: Binders}, we provide a considerably improved estimate of $S^{\textrm{sat.}}$ for the kagome lattice using Binder's 
algorithm. With which we may exactly calculate the partition function of a system of over, in our case, 1300 Ising spins at the 
saturation field with the expenditure of modest computational resources. 
We point out that it is only for the Ising kagome lattice, among other two dimensional lattices, that the zero field entropy 
exceeds the saturation field entropy.\\ 
The antiferromagnetic Ising models we investigate are described by the Hamiltonian
\begin{equation}
\label{eq: eq1}
 \mathcal{H} = \sum_{<i,j>}\sigma_{i}\sigma_{j} - h_c\sum_{i}\sigma_{i},
\end{equation}
on an $N$ site lattice with $|h_c| = z$, the nearest number of neighbors. This is the saturation field beyond which ordering sets 
in. The variables $\sigma _i = \pm 1$ which we represent by down and up spins, and the interaction between nearest neighbors is 
denoted by the angular brackets, setting the energy scale of the problem.
The boundary conditions are chosen to be either free or periodic. Although the number of allowed states for a given finite system 
will differ depending on the boundary conditions, the dominant multiplicative degeneracy of the system as $N \rightarrow \infty$ 
will reflect the bulk property.
It will then be a question of computational convenience whether free or periodic boundary conditions be chosen.
\begin{table*}[th!]
\caption{\label{Kag-chain}Partition function in configuration space $\mathcal{C}_m$ for $m-$site (cell) Ising chain (ladder $L_3$) 
with periodic and free boundary conditions; values for $m = 1, 2, 3, 4$ (which exclude the boundary spins/cells for free boundaries) are indicated. }
\begin{ruledtabular}
\begin{tabular}{lrr}
Boundary & Ising chain & Ladder $L_3$ \\
\hline
Periodic & $(\frac{1+\sqrt{5}}{2})^m + (\frac{1 - \sqrt{5}}{2})^m = 1, 3, 4, 7, \ldots$ 
& $(2 + \sqrt{3})^m + (2 - \sqrt{3})^m = 4, 14, 52, 194, \ldots$\\
Free & $\frac{1}{\sqrt{5}}\left[(\frac{1+\sqrt{5}}{2})^{m+2} - (\frac{1 - \sqrt{5}}{2})^{m+2} \right] = 2, 3, 5, 8, \ldots$ 
& $\frac{1}{2\sqrt{3}}\left[(2 + \sqrt{3})^{m+1} - (2 - \sqrt{3})^{m+1}\right] = 4, 15, 56, 209, \ldots$\\
\end{tabular}
\end{ruledtabular}
\end{table*}
\section{Ladders}
\label{sec: ladders}
\begin{figure}[bbp]
\centering
\includegraphics*[width=8.cm]{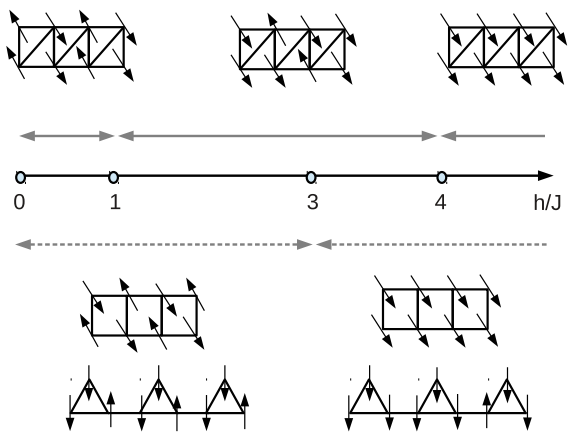}
\caption{Spin orderings on the Ising ladders $L_1, L_2$ (with spins tilted for clarity), and $L_3$ below and above the saturation fields; ordering 
regimes are indicated by gray full ($L_2$) and dashed ($L_1, L_3$) arrows. Trivial degeneracies due to translations are not indicated.}
\label{Orderings}
\end{figure}
In this section we describe the transfer matrix procedure to calculate the partition function $\Omega _m$ of a ladder with $m$ 
unit cells, such that no up spin may neighbor another; the space of such configurations is denoted as $\mathcal{C}_m$ which 
comprise the degenerate ground states at $h_c$ for $L_1$ and $L_2$; these two ladders are used primarily as (a) testbeds for our computations, and (b) for 
justifying the bounds in \eqref{eq: eq4} for the kagome lattice. We will demonstrate the idea with the case of ladder $L_3$ where a 
unit cell is taken to be a simple triangle; this will be relevant while building up the kagome lattice from this ladder. 
 We emphasize that, for this particular ladder $L_3$, the configurations in $\mathcal{C}_m$ do not constitute the degenerate space 
 of states at its critical field as justified in the succeeding paragraph. \\
 Below the saturation field value, the various phases on the ladders may be constructed as follows. For ladder $L_1$, clearly the up-down-up-down 
 configuration on each leg avoids any frustration at field strengths $|h| \leq z = 3$ and has energy per site $\epsilon^{L_1}_0 = -3/2$. 
 At $h = 3$, this energy equals that of the ferromagnetically ordered phase with energy per site $\epsilon^{L_1}_1 = 3/2 - h$, signaling a 
 first order transition. Similarly the down-down-up configuration for $L_3$, with the up spins along the base and energy 
 $\epsilon^{L_3}_0 = (-2-h)/3$, is the stable configuration at $h=0$; this phase persists up to $h = 3$ where it becomes degenerate with and 
 transits to the ferromagnetically aligned phase with energy $\epsilon^{L_3}_1 = 4/3 - h$. 
 We point out that the down-down-up configuration with the up spins
 along the apices (where $z = 2$) will be degenerate with the ferromagnetic phase at $h = 2$ but $\epsilon^{L_3}_0 = (-2-h)/3$ is more 
 stable at this phase point and therefore only a single transition exists for $L_3$. For ladder $L_2$, the three relevant phases and transition points may be readily 
 evaluated \cite{Morita} from Fig. 1 in Ref. [16] along the $J' = J$ line in the phase diagram of the ANNI chain. 
 These results of all three ladders are illustrated in Fig. \ref{Orderings}.\\
To illustrate the procedure, consider the configurations in $\mathcal{C}_m$ of ladder $L_3$; this calculation is relevant 
for sections \ref{sec: kagome} and \ref{sec: Binders} because it corresponds to an isolated ladder 
at the saturation field of the kagome lattice. At the saturation field of $L_3$, the configurations are different from $\mathcal{C}_m$ and lesser in number
because, as explained in the preceding paragraph, spin flips along the apices are not degenerate at the $h = 3$ transition.
However, the same procedure may be adopted for $S^{\textrm{sat.}}$ calculations for all three ladders.\\
The transfer matrix procedure for configuration space $\mathcal{C}_m$ in ladder $L_3$ is as follows: $L_3$ may be thought of as a simple linear chain, with the provisos 
that (a) now 4 states are permitted on each 'site' i.e. 
$\mathcal{C}_1 = $\{
\begin{tikzpicture}[scale=3]
 \draw(0,0) -- (1*0.2,0);
\draw (0,0) -- (0.5*0.2,0.866*0.2); 
 \draw (0.5*0.2,0.866*0.2) -- (1*0.2,0);
 \draw[<-](0.5*0.2,0.866*0.2 - 0.08) -- (0.5*0.2,0.866*0.2+1*0.05);
 \draw[<-](0.0,-0.05) -- (0,1*0.1);
  \draw[<-](1*0.2,-0.05) -- (1*0.2,0.1);
\end{tikzpicture}\tabf, \tabf
\begin{tikzpicture}[scale=3]
 \draw(0,0) -- (1*0.2,0);
\draw (0,0) -- (0.5*0.2,0.866*0.2); 
 \draw (0.5*0.2,0.866*0.2) -- (1*0.2,0);
 \draw[<-](0.5*0.2,0.866*0.2 - 0.08) -- (0.5*0.2,0.866*0.2+1*0.05);
 \draw[<-](0.0,-0.05) -- (0,1*0.1);
  \draw[->](1*0.2,-0.05) -- (1*0.2,0.1);
\end{tikzpicture}\tabf, \tabf
\begin{tikzpicture}[scale=3]
 \draw(0,0) -- (1*0.2,0);
\draw (0,0) -- (0.5*0.2,0.866*0.2); 
 \draw (0.5*0.2,0.866*0.2) -- (1*0.2,0);
 \draw[<-](0.5*0.2,0.866*0.2 - 0.08) -- (0.5*0.2,0.866*0.2+1*0.05);
 \draw[->](0.0,-0.05) -- (0,1*0.1);
  \draw[<-](1*0.2,-0.05) -- (1*0.2,0.1);
\end{tikzpicture}\tabf, \tabf
\begin{tikzpicture}[scale=3]
 \draw(0,0) -- (1*0.2,0);
\draw (0,0) -- (0.5*0.2,0.866*0.2); 
 \draw (0.5*0.2,0.866*0.2) -- (1*0.2,0);
 \draw[->](0.5*0.2,0.866*0.2 - 0.08) -- (0.5*0.2,0.866*0.2+1*0.05);
 \draw[<-](0.0,-0.05) -- (0,1*0.1);
  \draw[<-](1*0.2,-0.05) -- (1*0.2,0.1);
\end{tikzpicture}\tabf \},
and (b) the third of these states may not follow the second of these states on the chain. Following Metcalf and Yang 
\cite{Metcalf}, the transfer matrix for the present case may be defined as
\[
 M_{L_3} = \left( \begin{array}{cccc}
1&1&1&1\\
1&1&0&1\\
1&1&1&1\\
1&1&1&1
 \end{array} \right),
\]
where a zero entry indicates the aforementioned disallowed sequence of states. Under periodic boundary conditions, the total 
number of states is given by the trace of the transfer matrix over the $m$ cells \cite{Metcalf}. That is
\begin{equation}
\label{eq: eq2}
 \Omega ^{\textrm{PBC}} _m = \textrm{Tr}[(M_{L_3})^m] = (2+\sqrt{3})^m + (2-\sqrt{3})^m.
\end{equation}
Note that the trace automatically disallows the reverse of condition (b) i.e. $\mathcal{C}_1(2)$ not following $\mathcal{C}_1(3)$ 
through the chain ends; thus the non-Hermiticity of $M_{L_3}$ poses no issues. \\
We treat the boundary conditions on a more general footing by solving for the characteristic polynomial of $M_{L_3}$ to give 
$\lambda ^2(\lambda ^2 -4\lambda + 1) = 0$, from which the difference equation relating the partition function $\Omega _m$ of 
$m-$cell ladders may be readily read off as
\begin{equation}
\label{eq: eq3}
 \Omega _m = 4\Omega _{m-1} - \Omega _{m-2},
\end{equation}
for both periodic and free boundary conditions, for each of which we merely have to set different initial conditions in \eqref{eq: eq3}. 
The partition functions of $\mathcal{C}_m$ with both boundary conditions are compared with that of the Ising linear chain at 
saturation in table \ref{Kag-chain}.\\
The entropy per cell is then given by the logarithm of the dominant contribution to $\Omega _m$. We may follow this procedure to obtain the 
saturation entropies of all the illustrated ladders in Fig. \ref{ladders}. The values and the generating difference equations are 
tabulated in table \ref{entropies}. The saturation field entropy of ladder $L_2$ checks with an earlier calculation on the ANNI chain \cite{Sela}. 
The addition of diagonal bonds, in proceeding from $L_1$ to $L_2$, clearly reduces the residual entropy associated per lattice site. \\
\section{Kagome lattice}
\label{sec: kagome}
\begin{figure*}[t]
\centering
\subfigure[ ]{\label{PBCvsFBC}
\includegraphics[scale=0.32]{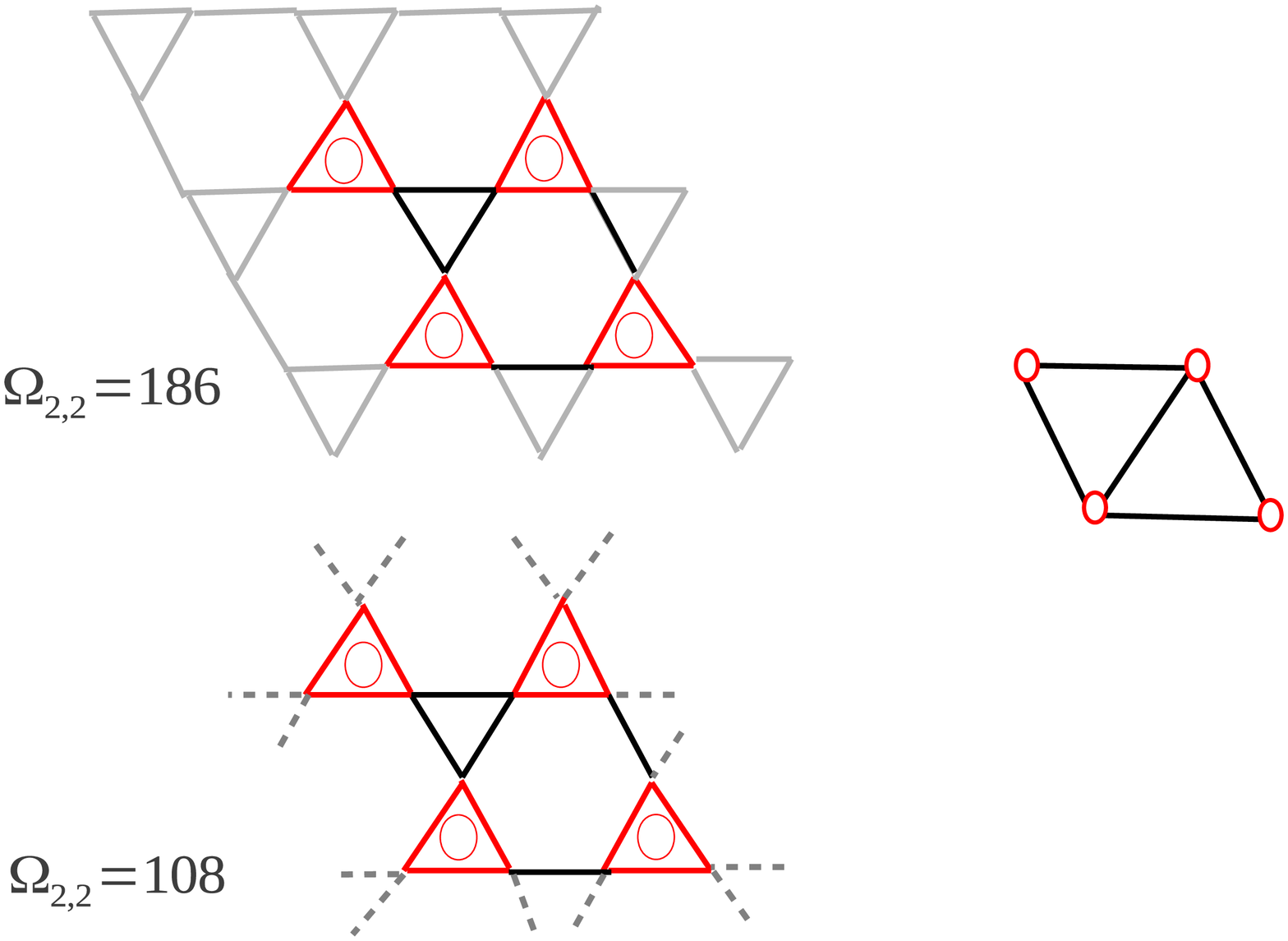}
}
\subfigure[ ]{\label{EntropyScalingPap}
\includegraphics[scale=0.32]{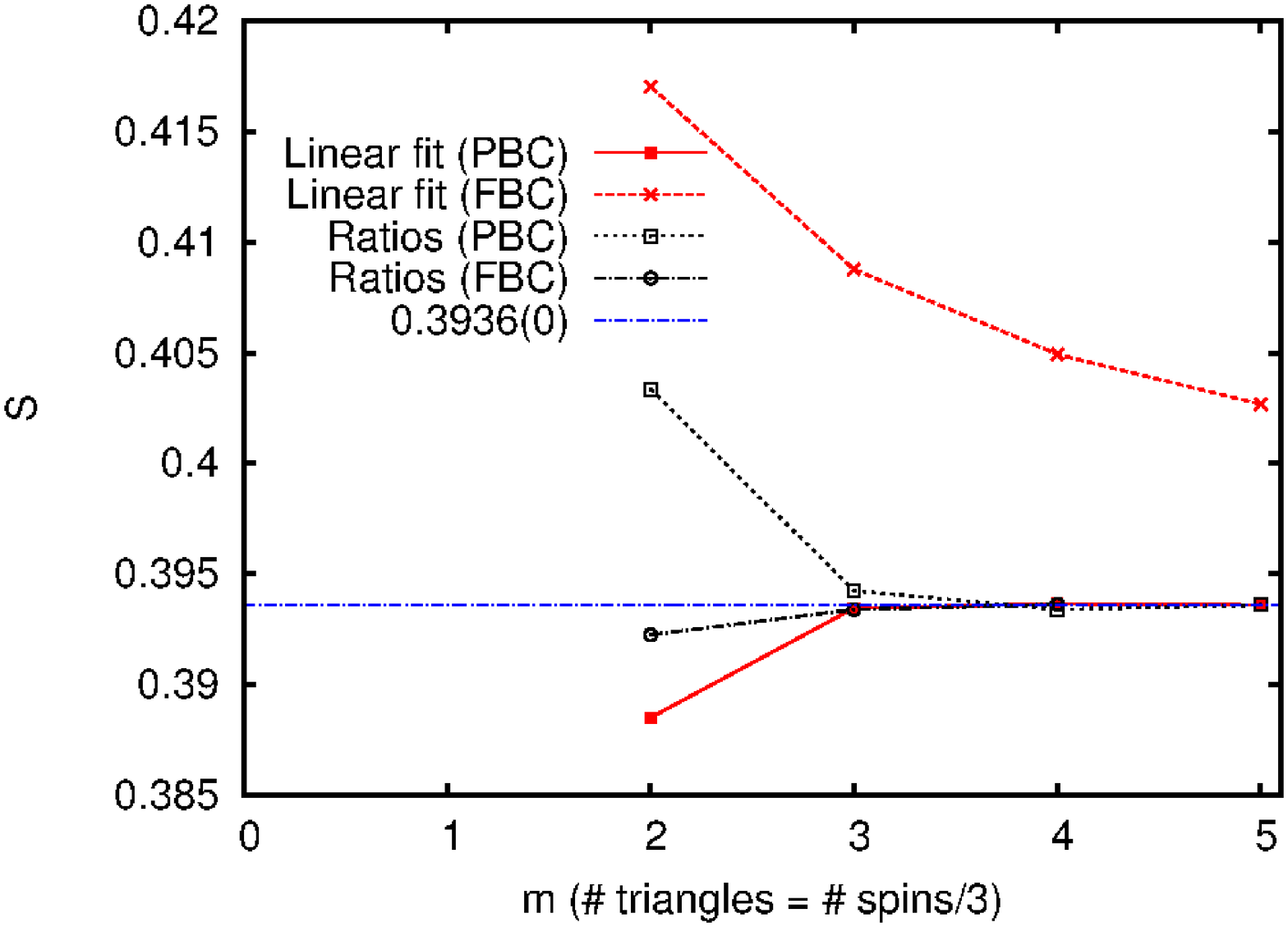}
}
\caption{(a) Free and periodic boundary conditions for the kagome lattice using an $m \times n = 2 \times 2$ system. Upper figure 
shows FBC: gray (light) triangles indicate spins aligned with the field, red (circled) triangles have finite degeneracies, black 
(dark) lines indicate the bonds constituting the triangular lattice. Bottom figure shows PBC with gray (dashed) lines indicating 
the imposition of periodicity. The partition function for each case and the equivalent triangular lattice are also indicated. 
(b) Scaling of residual saturation entropy, in units of $k_B$, on the Ising kagome lattice as function of number of triangles for 
two different scaling and boundary conditions. $m$ denotes the number of unit cell triangles on each ladder, and the number of such 
ladders $n = 100$.}
\end{figure*}
The kagome lattice, a section of which is illustrated in Fig. \ref{kagome}, may be thought of as ladder $L_3$ repeated in a two dimensional 
triangular lattice pattern, 
with a 'site' now being a simple triangle labelled $A, B, C$ in the figure. At zero field, the kagome Ising model is disordered 
\cite{Kano} while for fields below the onset of ferromagnetism, there is a finite net moment \cite{MoessnerSondhi}.
In these two regimes, the residual entropies for the Ising kagome are known accurately \cite{Kano, Udagawa, MoessnerSondhi}. And above the saturation field, 
the phase is ordered ferromagnetically.\\
At the saturation field, before proceeding with the calculations, we can provide 
upper and lower bounds for the kagome lattice's entropy at the very outset. For the lower bound, following the arguments in Ref. [9],
there must be more entropy per site than the triangular lattice because the increased connectivity of the latter serves to restrict 
the configuration space; we have already seen the reduction in entropy, while constructing $L_2$ from $L_1$, from the addition of 
diagonal bonds.
As regard an upper bound, following similar reasoning, clearly the kagome lattice cannot support more configurations than the 
ladder $L_3$ from which it is built. Therefore we get the inequality
\begin{equation}
\label{eq: eq4}
 0.3332427 \ldots < S_{\textrm{kag.}} < 0.4389859 \dots
\end{equation}
where the lower bound, the saturation entropy for the Ising triangular lattice, is known exactly through the solution of the 
hard-hexagon model \cite{Baxter}. The upper bound is obtained from the leading contribution to \eqref{eq: eq2} i.e. $\frac{1}{3}\log{(2 + \sqrt{3})}$,
with the result divided by $3$ because we consider the entropy per site in \eqref{eq: eq4}.\\
We adopt 2 approaches for estimating the convergence of the entropy as a function of system size. The first follows the transfer 
matrix and linear scaling method of Metcalf and Yang \cite{Metcalf}, for which we also provide an alternative reformulation; and 
the second is the ratios method of Milo\v{s}evi\'{c} et al. \cite{Milosevic}\\
In Fig. \ref{PBCvsFBC}, we illustrate how free and periodic boundary conditions are effected for an $m \times n = 2 \times 2$ 
kagome system. The black (dark) bonds indicate the underlying equivalent triangular lattice; this transformation to a triangular 
lattice makes the remainder of the analysis tractable.
\subsection{Transfer matrix: linear scaling}
For a two dimensional lattice the transfer matrices are constructed as follows from the one dimensional building chains 
\cite{Metcalf}, which in our case are the $L_3$ ladders. The matrix element $M_{i,j}$ is set to $0$ if the state $j$ of an 
$m$-cell ladder cannot follow state $i$ on an adjacent $m$-cell ladder; otherwise the matrix element is $1$. Clearly the matrix 
is of size $\Omega _{m} \times \Omega _{m}$, which already for $m=6$ gives a little over 7 million matrix elements in $M$.
The partition function is then given as before by $\Omega _{m,n} = \textrm{Tr}\left[M^n\right]$ for the $m \times n$ system; as 
we will see, typically $n = 100$ gives a good estimate up to three to four decimal places for the entropy. To obtain the entropy 
per $m$-cells, it is assumed that every new ladder added to the finite system multiplies the system's degeneracy by a constant 
factor of $\alpha$, so that
\begin{equation}
\label{eq: eq5}
 \log{\Omega _{m,n}} = n\log{\alpha} + C_{m,n},
\end{equation}
gives the entropy per $m$ cells as $\log{\alpha}$, where the $C_{m,n}$ denote the correction terms. It is expected that these 
terms decrease for increasing $m, n$ values. Thus the procedure is to calculate $\Omega _{m,n}$ and use the linear fit against 
$n$ to extract the entropy. We show in Fig. \ref{EntropyScalingPap} with full and dashed red lines the convergence of the entropy as the number of 
triangles is varied for periodic and free boundary conditions. Note that the trace operation automatically imposes periodic 
boundary conditions along the $n$-direction. Moreover we have checked for a system with free boundary conditions along the 
$n$-direction as well (using Binder's algorithm in section \ref{sec: Binders}) that the values obtained, and hence the convergence trends, 
are essentially the same. 
And as observed in Ref. [14] for other lattices, free boundary conditions does not give rapid convergence using \eqref{eq: eq5}.\\
As noted in the previous section (see \eqref{eq: eq3} or table \ref{entropies} for instance), the degeneracies may also be generated by solving 
difference equations on the lattice subject to appropriate initial values. For the kagome lattice, a difference equation for 
each $m$ is obtained and solved to obtain identical results as in Fig. \ref{EntropyScalingPap}. However this alternative and 
equivalent approach to Metcalf and Yang's procedure of matrix multiplication followed by the trace operation retains, at the 
present time, no computational gain because determining the characteristic polynomial of a matrix is about as hard as matrix 
multiplication with today's algorithms \cite{Keller-Gehrig}.
\subsection{Transfer matrix: ratios}
In the ratio method, the correction terms $C_{m,n}$ are substantially reduced by using a sequence of estimators for the entropy 
as \cite{Milosevic}
\begin{equation}
\label{eq: eq6}
 S_{m,n} = \log{\left[(\frac{\Omega _{m+1,n+1}}{\Omega _{m+1,n}})(\frac{\Omega _{m,n}}{\Omega _{m,n+1}})\right]}.
\end{equation}
For relatively large $m$ and $n$ values each added chain will multiply the system's degeneracy by a factor of 
$\alpha = \beta ^{3m}$, where $\beta$ is the factor associated with each site. Thus \eqref{eq: eq6} is seen to give the residual entropy 
per cell with considerable diminution of the correction terms. \\
As plotted in Fig. \ref{EntropyScalingPap} with the dotted and dashed-dotted black lines, the use of \eqref{eq: eq6} provides faster 
convergence for the entropy compared to \eqref{eq: eq5}; in contrast to \eqref{eq: eq5}, \eqref{eq: eq6} seems better suited for free boundary conditions. 
Also shown in the figure is the value of the estimator $S_{5,100}/k_B = 0.39360$ obtained from \eqref{eq: eq6} with free boundary conditions 
(using Binder's algorithm in section \ref{sec: Binders}), which differs from the $(\log{\alpha})_{5,100}$ value obtained from \eqref{eq: eq5} with periodic 
boundary conditions by approximately $0.00001$, thus giving three certain decimal places with an uncertainty in the fourth.\\
\begin{table*}[th!]
\caption{\label{entropies}Residual entropies per site, in units of $k_B$, at the saturation fields for the 
lattices in Fig. \ref{lattice}. The difference equations for the ladders are independent of the boundary conditions but the total number 
of states changes for each \textit{finite} segment.}
\setlength{\tabcolsep}{0.5em}
\begin{ruledtabular}
\begin{tabular}{ccr}
 Lattice&Difference equation &Entropy\\ \hline
 $L_1$&$x_n = 7x_{n-1} - 7x_{n-2} + x_{n-3}$&$\frac{1}{4}\log{(3 + 2\sqrt{2})} = 0.440686 \ldots$\\
$L_2$&$x_n = 5x_{n-1} - 2x_{n-2} + x_{n-3}$&$\frac{1}{4}\log{\left[5+ (\frac{187 - 9\sqrt{93}}{2})^{1/3} + 
(\frac{187 + 9\sqrt{93}}{2})^{1/3}\right]} -\frac{1}{4}\log{3} = 0.382245 \ldots$\\
$L_3$&$x_n = 3x_{n-1} - x_{n-2}$&$\frac{1}{3}\log{\left [(3 + \sqrt{5})/2\right ]} = 0.320807 \ldots$\\
 Kagome & - & 0.393589(6)
\end{tabular}
\end{ruledtabular}
\end{table*}
\section{Binder's algorithm}
\label{sec: Binders}
We have seen in the preceding section that free boundary conditions along with \eqref{eq: eq6} provide a rapidly convergent sequence for 
the entropy. 
The main limitation was however the calculation of $\Omega _{m,n}$ for large $\{m,n\}$ values. This may be achieved by employing 
Binder's algorithm towards an exact evaluation of the partition function of finite lattice systems \cite{Binder}. To briefly 
recapitulate, the partition function of a system of size $\{m, n\}$ is expressed in terms of the degeneracies $\gamma _{m,n}(i)$ 
of the $n^{\textrm{th}}$ ladder in its $i^{\textrm{th}}$ state. Then clearly
\begin{equation}
\label{eq: eq7}
 \Omega _{m,n} = \sum_i\gamma _{m,n}(i).
\end{equation}
Now the degeneracies of an added ladder for the $\{n, m+1\}$ system may be recursively computed by
\begin{equation}
\label{eq: eq8}
 \gamma _{m,n+1}(i) = \sum _{i'}\gamma _{m,n}(i'),
\end{equation}
with the summation running over only those values of $i'$ such that state $i$ may be adjacent to it. With this, we have computed 
the partition function of over 1300 spins with modest computational effort. For instance, we are able to reproduce up to 10 digits 
in the residual saturation field entropy value for the square lattice\cite{Milosevic} using twenty 10-rung $L_1$ ladders.\\
Using \eqref{eq: eq6} - \eqref{eq: eq8} we compute $S_{6,50}$, $S_{7,50}$ and $S_{8,50}$ to give six stable digits for the kagome lattice saturation field 
entropy
\begin{equation}
\label{eq: binder}
 S_{\textrm{kag}}/k_B = 0.393589(6).
\end{equation}
We compare this with low temperature Monte Carlo simulation results and the Bethe approximation for pyrochlore spin ice which, 
at the saturation field, may be described by a two-dimensional network of Ising pseudo-spin kagome lattice \cite{Isakov}. Scaling 
the saturation field results of Isakov et al. by a factor of 4/3 (because the corner spin in the pyrochlore tetrahedron is considered
frozen giving a high temperature entropy per site of only $\frac{3}{4}\log{2}$), we obtain the relevant results to be 
\begin{eqnarray}
\label{eq: Isakov}
 S_{\textrm{kag.}}^{\textrm{MC}}/k_B (T/J = 0.15) \approx 0.397, \nonumber \\
  S_{\textrm{kag.}}^{\textrm{Bethe}}/k_B \approx 0.38772.
\end{eqnarray}
This value is to be compared with the experimentally observed peak in the pyrochlore compound Dy$_{2}$Ti$_{2}$O$_{7}$ close to the 
high-field termination of the plateau \cite{Hiroi} where the physics was argued to be governed by decoupled kagome planes 
\cite{Isakov}. We point out, along with the authors of Ref. [7], that the computed values (in \eqref{eq: binder} and 
\eqref{eq: Isakov}) are slightly higher than the magnitude of the experimentally measured peak. 
For the spin ice compound at saturation, the corresponding value of approximately $2.4544$ 
Joules/deg./mole (obtained by multiplying \eqref{eq: binder} by $3R/4$, where $R$ is the gas constant) may be compared with the first
prominent experimental peak in the saturation entropy at $T = 1K$ of about $2.1$ Joules/deg./mole \cite{Hiroi}.
The difference between the two values suggests that either more precise measurements with error bars are required on this compound 
close to the transition field, or that the applicability of the Ising model on decoupled kagome planes close to this spin-ice material's 
saturation field might need to be slightly reconsidered.
\section{Summary}
We have considered the degenerate space of states of a few Ising ladders and the Ising kagome lattice at the saturation external 
field. For the ladders, by a simple redefinition of a site, the residual entropy may be exactly computed. We treat the generation 
of states for periodic and free boundary conditions on a general footing by use of difference equations.\\
For the kagome lattice, we are able to provide six stable digits for the residual entropy by calculating the exact partition 
function of over 1300 spins using Binder's algorithm implemented on a standard computer. Our accurate result compares reasonably 
with approximate results from low temperature Monte Carlo simulations and the Bethe approximation for an equivalent system. 
We believe that by constructing appropriate ladders the residual saturation field entropies of geometrically complex lattices, 
like the pyrochlore, may be similarly computed with more ease after their transformation to standard lattice structures. 
Comments on the relation to the experimental situation on the spin-ice compound Dy$_{2}$Ti$_{2}$O$_{7}$ were made.\\
The primary results are summarized in table \ref{entropies}.
\acknowledgments
The author thanks Oleg Derzhko, Sergei Isakov and Hartmut Monien for discussions, the latter for comments on the manuscript. 
Support of the Bonn-Cologne Graduate School through the Deutsche Forschungsgemeinschaft is gratefully acknowledged.

\end{document}